\documentclass[preprint]{revtex4-1}
\pdfoutput=1
\usepackage[utf8]{inputenc}
\usepackage[english]{babel}
\usepackage{amssymb,amsfonts,amsmath}
\usepackage{hyperref}

\usepackage[stretch=10]{microtype} 

\hypersetup{
    bookmarks=true,
    unicode=false,
    pdftoolbar=true,
    pdfmenubar=true,
    pdffitwindow=false,
    pdfstartview={FitH},
    pdftitle={Electrical Resistivity Model for Quasi-one-dimensional structures},
    pdfauthor={Stepan Botman},
    pdfsubject={Theoretical physics},
    pdfkeywords={resistivity} {Boltzmann equation} {quasi-one-dimensional},
    colorlinks=true,
    linkcolor=blue,
    citecolor=green,
}

\usepackage{graphicx}

\begin{document}

\title{Electrical Resistivity Model for Quasi-one-dimensional structures.}

\author{Stepan Botman}
\email{sbotman@innopark.kantiana.ru}
\affiliation{Institute of Chemistry and Biology, Immanuel Kant Baltic Federal University}

\author{Sergey Leble}
\email{leble@mif.pg.gda.pl}
\affiliation{Institute of Physics and Technology, Immanuel Kant Baltic Federal University}
\date{\today}

\begin{abstract}
In this paper electron-impurity scattering coefficient of Bloch waves for one dimensional Dirac comb potential is used for calculation of temperature dependence of resistivity within kinetic theory. We restrict ourselves by scattering on impurities that is also modelled by zero-range potential. The standard averaging is expressed by integral that is evaluated within advanced numerical procedure. The plots  on base of the calculations results demonstrate strong variability as function of temperature and impurity strength. 
\end{abstract}


\maketitle

\section{Introduction}

Quasi one-dimensional structures such as nanowires, nanorods and nanoribbons are important due to wide range of application, particularly as photodetectors, logic devices, thermoelectric coolers, chemical and biological sensors etc.
Knowledge of electronic properties this class of materials is essential for practical application.

Latest theoretical and experimental results in the field of transport in low-dimensional structures have demonstrated that by changing particular conditions and parameters of such  objects, the properties of conductivity in low-dimensional materials can be drastically changed from highly conductive to the insulator regime. 
The transport properties of low dimensional systems is of intense interest for the physics of condensed matter and has been the subject of investigation by many physicists and some mathematicians for over half a century ~\cite{Dingle50}.
Recent progress raises both general academic questions and technological demands in understanding the transport phenomena through low-dimensional and nanosystems.

In order to elaborate an understanding in the challenging problem of transport in low-dimensional and nanosystems theoretical model should be created. Although various calculations methods may give reasonable results for conductivity in some cases, they do not insight into basics concepts of transport properties of this type of systems.

\subsection{Bismuth nanowires}

Study of Bi nanowires showed that it's resistance is non-trivially depend on temperature and diameter of samples~\cite{Zhang00}: for small diameter normalized resistance monotonically decreases with temperature, but for bigger diameters local maximum is formed. Although some semi-empirical models were created~\cite{lin2000theoretical}, this phenomena still needs more general theoretical consideration.
In pure single crystal Bi nanowires boundary scattering process is believed to be dominant. Thus, at low temperatures (up to 300~K) phonon contribution to resistivity is not crucial. One may conclude that scattering on defects determine general resistivity properties of the system.

Basic idea of the work is to create and verify general model for quasi 1D systems, revealing conduction dependence on temperature and structure parameters. 

\section{Resistivity calculation}
\subsection{1D model}

\begin{figure*}[tp]
\begin{minipage}[h]{0.5\linewidth}
\center
\centering
\includegraphics[width=\textwidth]{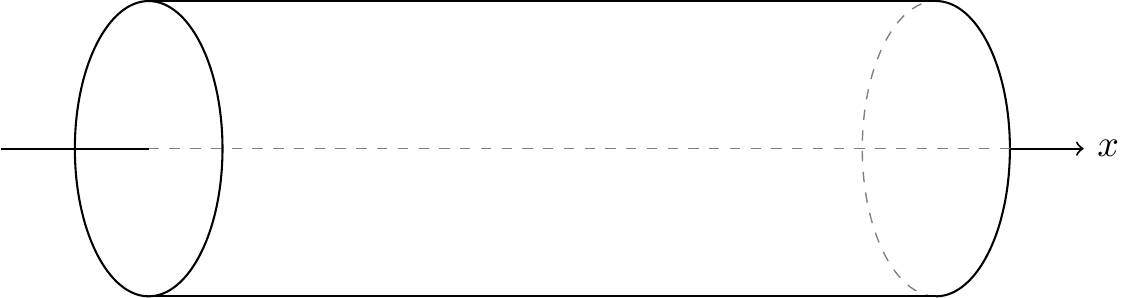}
\includegraphics[width=\textwidth]{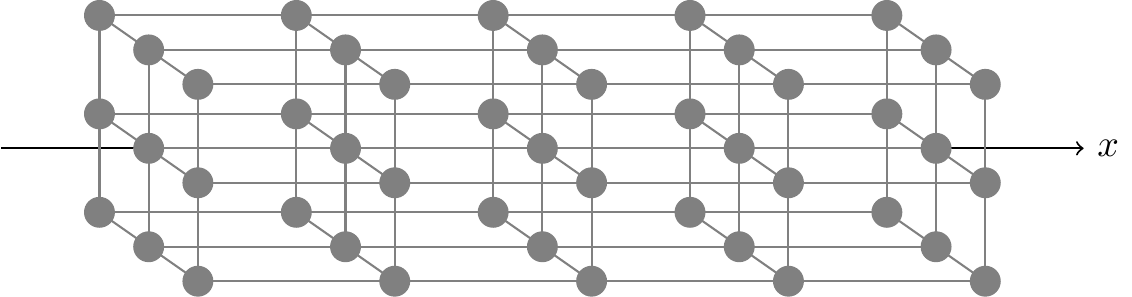}
\includegraphics[width=\textwidth]{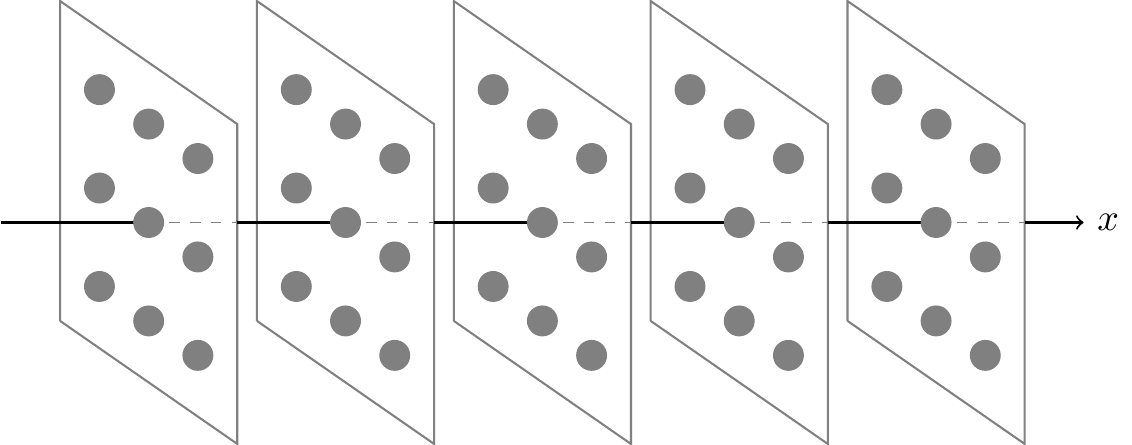}
\includegraphics[width=\textwidth]{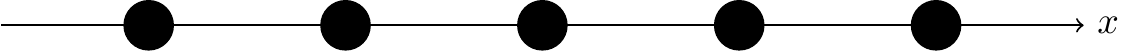}
\end{minipage}
\caption{Basic idea: averaging in transverse direction.}
\label{transversal-averaging}
\end{figure*}

The basic idea of model is applying some averaging procedure (exact form of averaging is not known), which reduce number of dimensions to one, as shown in picture~(\ref{transversal-averaging}).
Let us choose for a model electron moving in the Dirac comb potential --- i.e. potential of equidistant Dirac delta functions:
\begin{equation}
\hat{V} = \beta \delta (x-na),
\; n = 0, \pm 1, \ldots
\label{diraccomb-potential}
\end{equation}
where \(\beta\) --- parameter of potential, \(a\) --- period of cell.
Each term of~(\ref{diraccomb-potential}) represents transversal layer of atoms.

From previous results~\cite{TASK2016} we get analytical expressions for band structure and for several physical quantities such as density of states, electron velocity and effective mass:
\begin{gather}
K(E) = \frac{1}{a} \arccos \left[ 
\cos \left(\frac{\sqrt{2mE}}{\hbar} a \right) \right.
+ \left. \frac{m \beta}{\hbar \sqrt{2mE}} 
\sin \left(\frac{\sqrt{2mE}}{\hbar} a \right)
\right],
\label{energy-dispersion}
\\
\rho (E) = \frac{2}{\pi} \left( \frac{d E}{d K} \right)^{-1} = 
\frac{2}{\pi} K'(E), 
\label{dos-analytical}
\\
v (E) = \frac{1}{\hbar} \frac{d E}{d K} =
\frac{1}{\hbar} \frac{1}{K'(E)},
\label{velocity-analytical}
\\
m^* = \frac{1}{\hbar^2} \frac{d^2 K}{dE^2}.
\label{mass-analytical}
\end{gather}

\begin{figure*}[t]
\centering
\includegraphics[width=0.75\textwidth]{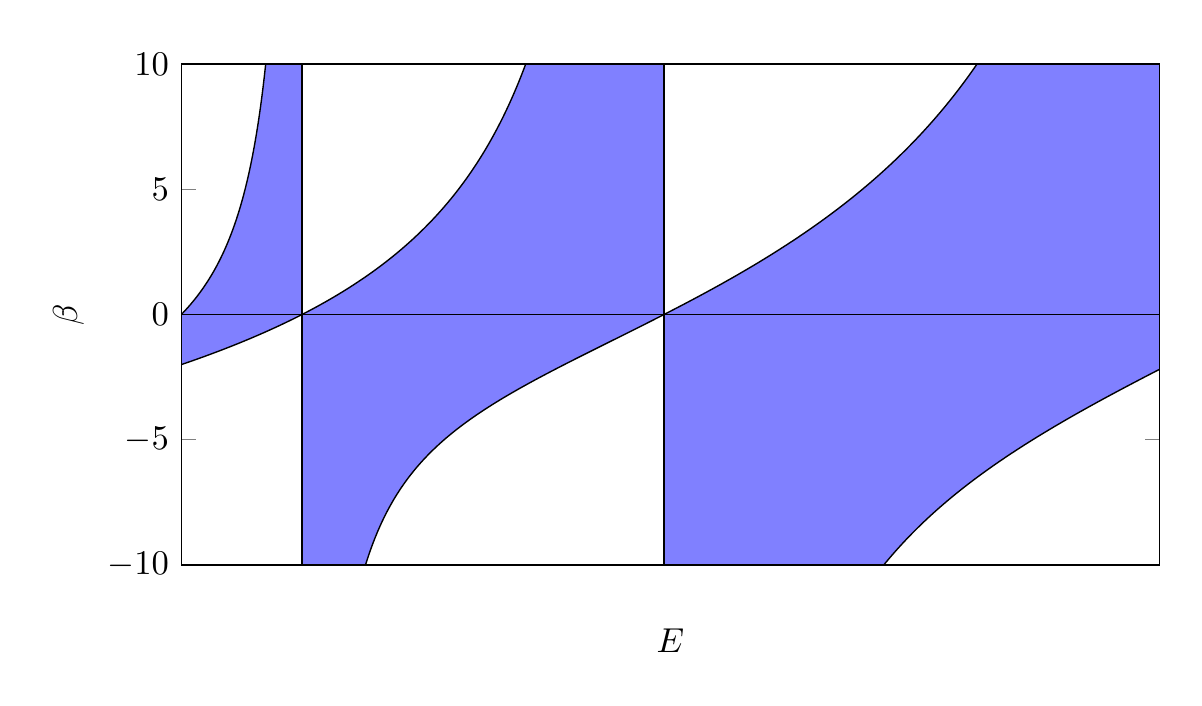}
\caption{Zone edges depending on \(\beta\) parameter.}
\label{edos}
\end{figure*}

Analytical form of~(\ref{dos-analytical}), (\ref{velocity-analytical}) and~(\ref{mass-analytical}) can be trivially obtained.

One may notice that~(\ref{dos-analytical}) can be easily integrated over \(E\) which allow straightforward calculation of Fermi energy.

\subsection{Calculation workflow}

\begin{figure}[p]
\centering
\includegraphics[width=\textwidth,height=\textheight,keepaspectratio]{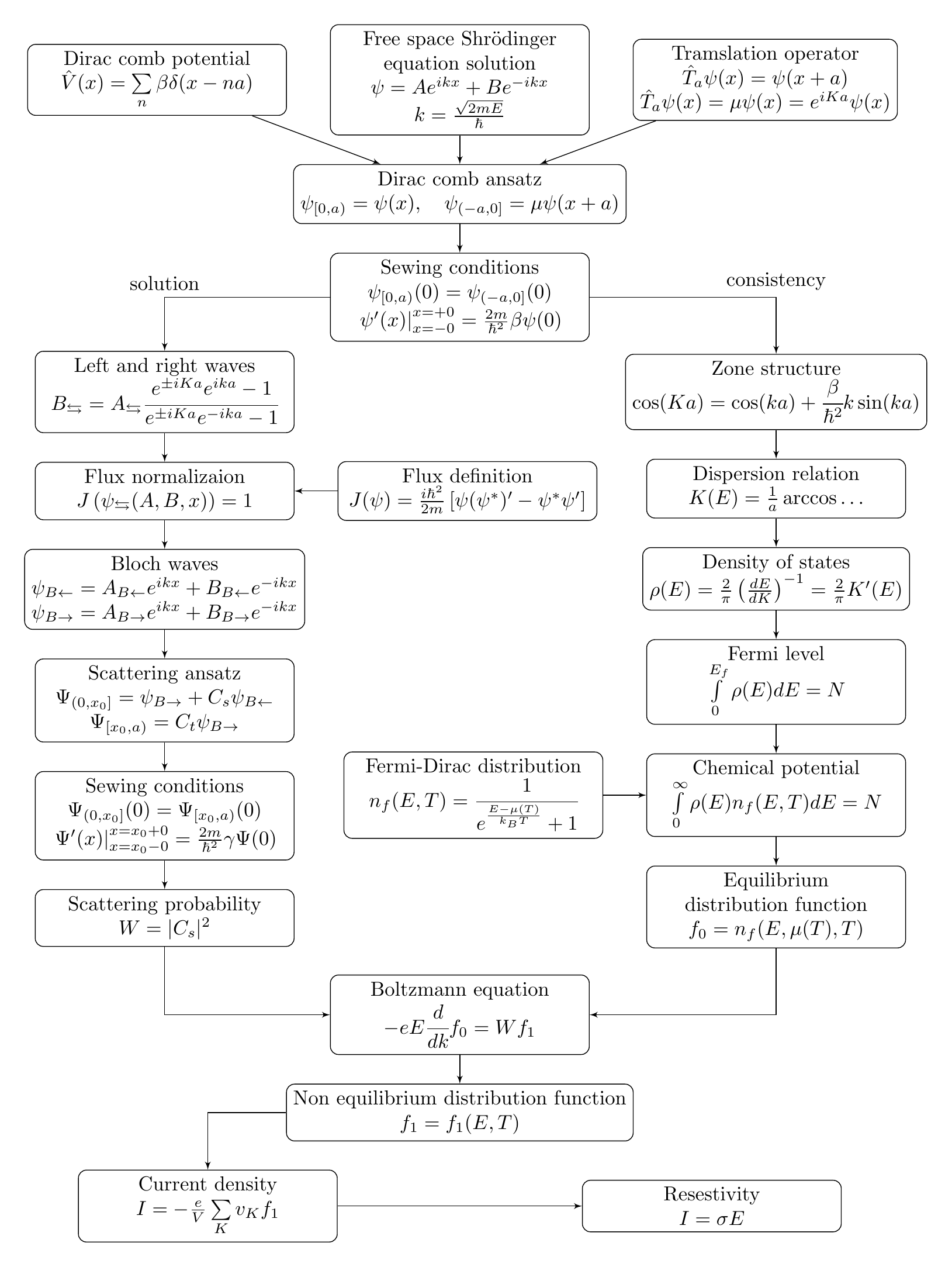}
\caption{Resistivity calculation workflow.}
\label{calculation-workflow}
\end{figure}

The calculation workflow is shown in figure~(\ref{calculation-workflow}). Using scattering probability and density of states, one can calculate resistivity for a system described above using kinetic equation.

\subsection{Kolmogorov equation}
Thus, in order to study conductivity of nanostructures It's more convenient to use Kolmogorov equation as defects scattering process prevalent for them.

A kinetic  equation describes evolution of a distribution function \(f \left( \boldsymbol k, \boldsymbol r, t \right)\) in phase space. There are two factors which cause \(f\) to evolve: scattering which causing discontinuous changing of \(\boldsymbol k\) (right side of equation) and drift and acceleration of the particles (left side of equation). 
Generally, Kolmogorov~\cite{Kolmogorov} equation can be written as:
\begin{equation}
\frac{\partial f}{\partial t} + \dot{\boldsymbol k} \frac{\partial f}{\partial \boldsymbol k} +
\dot{\boldsymbol r} \frac{\partial f}{\partial \boldsymbol r} = \left.
\frac{\partial f}{\partial t} \right|_{coll}.
\end{equation} 

Taking into account semi-classical equations of motions: \(\dot{\boldsymbol r} = \boldsymbol v\) and \(\hbar \dot{\boldsymbol k} = \boldsymbol F\), and assuming electro-magnetic field applied to a system \(\boldsymbol F = - e \left( \boldsymbol E + \frac{1}{c} \left[ \boldsymbol v \times \boldsymbol H \right] \right)\), Boltzmann equation can be written as:
\begin{equation}
\frac{\partial f}{\partial t} + \boldsymbol v
\frac{\partial}{\partial \boldsymbol r} f - e
\left( \boldsymbol E + \frac{1}{c}
\left[ \boldsymbol v \times \boldsymbol H \right] \right)
\frac{\partial}{\partial \boldsymbol k} f =
I \left( f \right),
\label{Boltzman}
\end{equation}
where \(f \left( \boldsymbol k, \boldsymbol r \right)\) is ensemble average nonequilibrium distribution function, \(I \left( f \right)\) is a collision integral. For scattering on static potential one may use:
\begin{equation}
I \left( f \right) = \sum \limits_{\boldsymbol k'}
W \left( \boldsymbol k, \boldsymbol k'\right)
\left[ f \left( \boldsymbol k' \right) - 
f \left( \boldsymbol k \right) \right],
\label{int-collision}
\end{equation}
where \(W \left( \boldsymbol k, \boldsymbol k'\right)\) is a collision probability.

Let's rewrite equation (\ref{Boltzman}) for electrons in solid making several simplifications. Firstly, we will assume that applied filed is electrostatic (\(\boldsymbol H = \boldsymbol 0\)). Secondly, we will focus on static conductivity problem (\(\frac{\partial f}{\partial t} = 0\)). Thirdly, we will adopt homogeneous current hypothesis (\(\frac{\partial f}{\partial r} = 0\)). Fourthly, we will assume that deviation of distribution function from its equilibrium is small (\(f=f_0+f_1\), \(\left| f_1 \right| \ll f_0\)). Because of equilibrium distribution function \(f_0\) is \(\boldsymbol k\)-symmetric, it has no impact in collision integral.

Applying all before-mentioned simplifications one can rewrite (\ref{Boltzman}) as follows:
\begin{equation}
- e \boldsymbol E \frac{\partial f_0}{\partial \boldsymbol k} 
= \sum \limits_{\boldsymbol k'} W \left( \boldsymbol k, \boldsymbol k'\right)
\left[ f_1 \left( \boldsymbol k' \right) - 
f_1 \left( \boldsymbol k \right) \right].
\label{Boltzman-fin}
\end{equation}

For one dimensional case (\( W \left( \boldsymbol k, \boldsymbol k'\right) = W(k,-k) = W\), \(f_1(-k)=-f_1(k)\)) further simplification is possible:
\begin{equation}
f_1 = \frac{1}{2 W} e \boldsymbol E \frac{\partial f_0}{\partial \boldsymbol k}.
\label{non-equlibrium-f}
\end{equation}

Solution of this equation gives us a non-equilibrium distribution function for a given external field \(\boldsymbol E\) and collision mechanism \(I \left( f \right)\), which in turn can be used to determine electric current (in form of electron quasiparticle propagation studied by Drude, Sommerfeld, Bloch, Landau).
\begin{equation}
\boldsymbol j = - \frac{e}{V} \sum \limits_k
\boldsymbol v_k f,
\label{current-simple}
\end{equation}
where summation is performed over qusiparticle states \(k\), \(\boldsymbol v_k\) is group velocity \(\boldsymbol v_k = \frac{1}{\hbar} \frac{\partial \varepsilon_k}{\partial\boldsymbol k}\), \(V\) is sample volume. For real systems It's useful to transform sum in~(\ref{current-simple})  into integral:
\begin{equation}
\boldsymbol j = -e n_e \frac{\hbar^2}{m^2} \int
\boldsymbol k f \! \left( k \right) d^3k.
\label{current-simple-int}
\end{equation}

As the result we can find resistivity of the system by comparing (\ref{current-simple-int}) with classical definition of resistivity:
\begin{equation}
\boldsymbol j = \sigma \boldsymbol E.
\label{resistivity}
\end{equation}

\subsection{Numerical integration}
\subsubsection{Quadrature formula}
Consider the following integral:
\begin{equation}
\int \limits_0^\infty F(E) n'_f (E, \mu, T) dE \approx
\int \limits_{E_{\min}}^{E_{\max}} F(E) n_f' (E,T,\mu) dE
\end{equation}

Fermi-Dirac distribution function:
\begin{equation}
n_{f} (E,T)= \frac{1}{e^{\frac{E-\mu(T)}{k_B T}} + 1},
\label{Fermi-distribution}
\end{equation}
where \(k_B\) --- Boltzman constant, \(\mu\) --- chemical potential ( \(\mu(T=0)=E_f\), \(E_f\) --- Fermi energy), \(T\)~---~temperature.

Behavior of functions \(K(E)\) and \(n'_F (E, \mu, T)\) makes it impossible to calculate integral with classical numerical integrations techniques. Thus, one should create appropriate quadrature formula.

In order to construct proper numerical procedure let's note that \(n_f'\) --- rapidly changing function, and \(F(E)\)  (which is \(K(E)\) or other function) --- slowly changing function.
\begin{multline}
\int \limits_{E_{\min}}^{E_{\max}} n_f' (E,T,\mu) F(E) dE
= \sum \limits_{i=1}^{N-1} \int \limits_{E_i}^{E_{i+1}}
n_f' (E,T,\mu) F(E) dE \approx \\
\approx \sum \limits_{i=1}^{N-1} F(\bar{E}_i)
\int \limits_{E_i}^{E_{i+1}} n_f' (E,T,\mu) dE
= \sum \limits_{i=1}^{N-1} F(\bar{E}_i) \left[ 
n_f (E_{i+1},T,\mu) - n_f (E_i,T,\mu) \right].
\label{numerical-proc}
\end{multline}

In equation~(\ref{numerical-proc}) we used the fact that function \(F(E)\) can be replaced for every small interval \(\left[E_i,E_{i+1}\right]\)  with it's value \(F(\bar{E}_i)\) at point \(\bar{E}_i \in \left[E_i,E_{i+1}\right]\), and taken outside the integral sign.

As was shown before, derivative of function \(K(E)\) equals infinity at band edge points. In this case, procedure~(\ref{numerical-proc}) will not suit and one should use trapezoidal method as fallback.

\begin{figure}
\begin{minipage}[h]{0.45\linewidth}
\center{
\includegraphics[width=\textwidth]{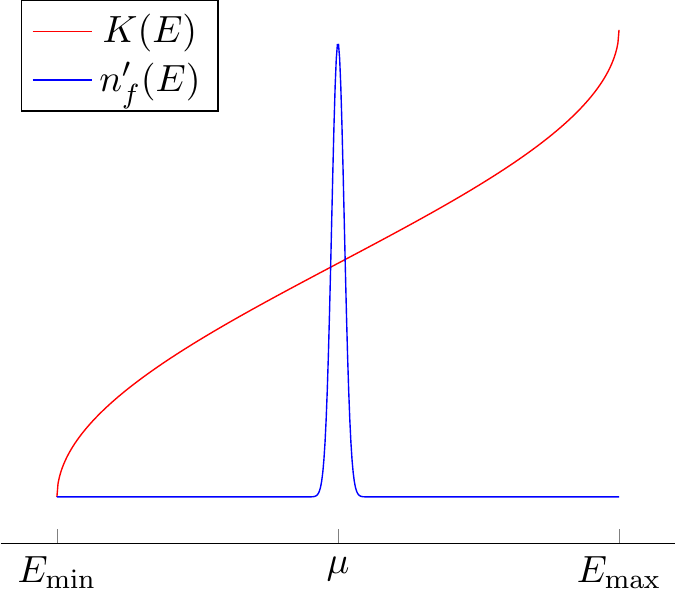}
\\ a) }
\end{minipage}
\begin{minipage}[h]{0.45\linewidth}
\center{
\includegraphics[width=\textwidth]{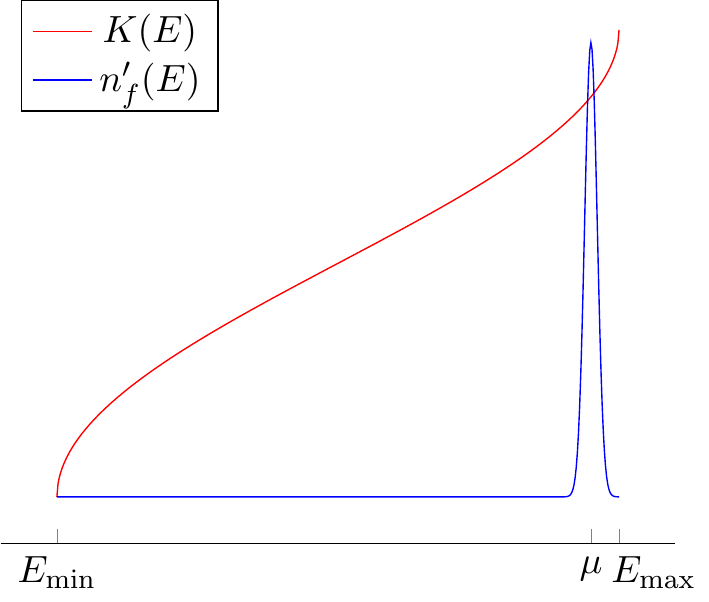}
\\ b) }
\end{minipage}
\caption{Mutual position of \(n'_f(E,\mu)\) (blue) and \(K(E)\) (red) functions.}
\end{figure}

\subsubsection{Grid}

The fact that \(n'_f(E)\) decays exponentially to zero on both sides of point \(E=\mu\) allows us to shrink integration area to a relatively small interval around the peak. Integration interval was set in terms of peak width \(2k_BT\). It's natural to set grid centered on peak of \(n'_F(E,\mu,T)\) functions.

In case if \(\mu\) is far from band edges, integration interval is \(\left[\mu-mk_BT,\mu+mk_BT\right]\), where \(m\) --- some positive number. In case if \(\mu\) is near band edge, grid chosen in the way that both \(\mu\) and nearest band edge are points of the grid.

\subsubsection{Error estimation}
Absolute error for~(\ref{numerical-proc}) can be estimated as follows:
\begin{equation}
\int \limits_{E_{\min}}^{E_{\max}} n_f' (E,T,\mu) F(E) dE
= \sum \limits_{i=1}^{N-1} F(\bar{E}_i)
\int \limits_{E_i}^{E_{i+1}} n_f' (E,T,\mu) dE + Q
\end{equation}
where
\begin{equation}
Q = \sum \limits_{i=1}^{N-1} R_i(E)
\int \limits_{E_i}^{E_{i+1}} n_f' (E,T,\mu) dE
\leq \frac{M h}{2} \int \limits_{E_1}^{E_2}
n'_f (E, \mu, T) dE \leq \frac{M h}{2}
\end{equation}
Next,
\begin{equation}
\left| R_i(E) \right| = \left| \left. \frac{dF}{dE} \right|_{E=\xi}
(E-\bar{E}_i) \right| \leq \frac{h}{2}
\max_{\left[E_{\min},E_{\max}\right]} \left| \frac{dF}{dE} \right|
= \frac{M h}{2}
\end{equation}
where \(\xi \in \left[E_i,E_{i+1}\right]\).

In order to estimate error, associated with integration interval reduction, let us consider the following integral:
\begin{equation}
\int \limits_{E_1}^{E_2} n'_f (E, \mu, T) dE =
\left. \vphantom{\int} n_f (E, \mu, T) \right|_{E_1}^{E_2} =
\left. \vphantom{\int} \frac{1}{\frac{E-\mu}{k_B T} +1} \right|_{E_1}^{E_2}.
\label{fdint}
\end{equation}

Let \(E_1=\mu - m k_B T\) and \(E_2=\mu + m k_B T\), where \(m\) --- some positive number.
Let us substitute \(E_1\) and \(E_2\) to \ref{fdint} and set it equal to full integral \(-1\) plus error associated with truncating \(\Delta\):
\begin{equation}
\left. \vphantom{\int} \frac{1}{\frac{E-\mu}{k_B T} +1} 
\right|_{\mu - m k_B T}^{\mu + m k_B T} =
\frac{1 - e^m}{e^m + 1} = -1 + \Delta
\end{equation}
Which means that if the integral of remainder parts equals \(\Delta\) and for a given \(\Delta\) one can write:
\begin{equation}
m = \ln \left(\frac{2}{\Delta} - 1 \right) 
\approx - \ln \left( \frac{\Delta}{2} \right).
\end{equation}
which means that:
\begin{equation}
\int \limits_{\mu - m k_B T}^{\mu + m k_B T} n'_f (E, \mu, T) dE = -1 + \Delta.
\end{equation}

\subsection{Current and resistivity calculations}

Here on should solve the Kolmogorov equation with appropriately constructed collision integral, which was described earlier.

Equation for total current:
\begin{equation}
J = e \int f_1 \frac{p}{m} dp = e \int f_1 v(E) \rho(E) dE = \frac{2 e}{\pi \hbar} \int f_1 dE.
\end{equation}

Obviously, current density is just \( j = J/V \).

Finally, combining (\ref{resistivity}) and (\ref{non-equlibrium-f}) one may write expression for resistivity:
\begin{equation}
\sigma = \frac{e^2}{\pi \hbar} \int \frac{1}{W} \frac{\partial f_0}{\partial \boldsymbol k} dE.
\end{equation}

Or, substituting \(W=|C_s|^2\):
\begin{equation}
\sigma = \frac{e^2}{\pi \hbar} \int
\left|
- \frac{1}{B_-}
\frac{ \left( b_+ e^{ik x_0} + e^{-ik x_0} \right)^2 \gamma m}
{( b_- - b_+ )( i \hbar^2 k + \gamma  m ) + \gamma m \left( b_- b_+ e^{2 ik x_0} + e^{-2 ik x_0} \right)}
\right|^2
\cdot \frac{\partial f_0}{\partial k} dE,
\end{equation}

Or:
\begin{equation}
\sigma = \frac{e^2}{\pi m} \int
 k \left( 1 - |b_-|^2 \right)
\left|
\frac{ \left( b_+ e^{ik x_0} + e^{-ik x_0} \right)^2 \gamma m}
{( b_- - b_+ )( i \hbar^2 k + \gamma  m ) + \gamma m \left( b_- b_+ e^{2 ik x_0} + e^{-2 ik x_0} \right)}
\right|^2
\cdot \frac{\partial f_0}{\partial k} dE,
\label{resitivity-int}
\end{equation}

Where radius dependence is: \(\gamma=\gamma(r)\), \(f_0=f_0(T,\mu(r))\). (but \(\beta=\text{const} \) )

In (\ref{resitivity-int}) and \(b_{\pm}\) stands for:
\begin{equation}
b_{\pm} = \frac{e^{\pm iKa}e^{-ika} -1}{e^{\pm iKa}e^{ika} -1}.
\end{equation}

Obviously, \(b_{\pm} = b_{\pm} (k, \beta) \).

Expression for \(K(E)\):
\begin{equation}
K(E) = \frac{1}{a} \arccos \left[ 
\cos \left(\frac{\sqrt{2mE}}{\hbar} a \right) + \right.\\
\left. \frac{m \beta}{\hbar \sqrt{2mE}} 
\sin \left(\frac{\sqrt{2mE}}{\hbar} a \right)
\right],
\end{equation}

Equilibrium distribution function is Fermi-Dirac distribution function~(\ref{Fermi-distribution}).

\subsection{Evaluation of integral for low temperatures.}

Let's consider integral in the following form:
\begin{equation}
I = \int \limits_0^\infty F(E) n'_f (E,T) dE,
\label{int-inital}
\end{equation}
where \(F(E)\) --- some function of energy, \(n'_f (E,T)\) --- first derivation of the Fermi-Dirac distribution function:
\begin{equation}
n_f (E,T) = \frac{1}{e^\frac{E-\mu}{k_B T} +1}.
\end{equation}

It's known, that at low temperatures \(n'_f (E,T)\) has a sharp peak at \(E=\mu\). Taking this into consideration, we replace lower integration limit with \(-\infty\) and expand \(F(E)\) to Taylor series at \(\mu\) point:
\begin{equation}
F(E) = \sum \limits_0^\infty \frac{F^{(n)}(\mu)}{n!} (E-\mu)^n.
\end{equation}

Thus, for integral we have:
\begin{equation}
I = \int \limits_0^\infty F(E) n'_f (E,T) dE =
\sum \limits_0^\infty \frac{F^{(n)}(\mu)}{n!} 
\int \limits_{-\infty}^\infty n'_f (E,T) (E-\mu)^n dE.
\end{equation}

Next, we introduce new variable: \(z=\frac{E-\mu}{k_B T}\):
\begin{equation}
I = - \sum \limits_0^\infty \frac{F^{(n)}(\mu)}{n!} 
\int \limits_{-\infty}^\infty \frac{e^z}{(e^z+1)^2} (zk_BT)^n dz.
\end{equation}

Obviously, \(e^z/(e^z+1)^2=1/2(\cosh(z)+1)\) is odd function and \((zk_BT)^n\) is either odd or even depending on number \(n\). Thus, integral can be transformed into:
\begin{equation}
\begin{split}
I &= - \sum \limits_0^\infty \frac{F^{(2n)}(\mu)}{2n!} 
\int \limits_{0}^\infty \frac{e^z}{(e^z+1)^2} (zk_BT)^{2n} dz \\
&= - F(\mu) - \sum \limits_1^\infty \frac{F^{(2n)}(\mu)(k_BT)^{2n}}{2n!} 
\int \limits_{0}^\infty \frac{e^z}{(e^z+1)^2} z^{2n} dz
\end{split}
\label{int-zerotemp-series}
\end{equation}

Equation~(\ref{int-zerotemp-series}) describes how integral~(\ref{int-inital}) behave at low temperatures. The first order approximation gives us a quadratic temperature dependence.

In order to achieve some numerical results we should calculate integral in~(\ref{int-zerotemp-series}). Let's start from integration by parts:
\begin{multline}
\int \limits_{0}^\infty \frac{e^z}{(e^z+1)^2} z^{2n} dz =
\underbrace{ \left. -\frac{z^{2n}}{e^z+1} \right|_0^\infty}_{= 0} +
\int \limits_{0}^\infty \frac{2n z^{2n-1}}{e^z+1} dz =
\int \limits_{0}^\infty 2n z^{2n-1} \frac{e^{-z}}{e^{-z}+1} dz =\\
=\int \limits_{0}^\infty 2n z^{2n-1} e^{-z} 
\left[ \sum \limits_{m=0}^{\infty} (-e^{-z})^m \right] dz =
\int \limits_{0}^\infty \sum \limits_{m=0}^{\infty}
(-1)^m z^{2n-1} e^{-(m+1)z} dz 
\stackrel{x=(m+1)z}{=\joinrel=\joinrel=\joinrel=\joinrel=} \\
=\int \limits_{0}^\infty \sum \limits_{m=0}^{\infty}
(-1)^m \left( \frac{x}{m+1}\right)^{2n-1} \frac{e^{-x}}{m+1} dx =
\sum \limits_{m=1}^{\infty} \frac{(-1)^{m+1}}{m^{2n}}
\int \limits_{0}^\infty x^{2n-1} e^{-x} dx = \\
= \sum \limits_{m=1}^{\infty} \frac{(-1)^{m+1}}{m^{2n}} \Gamma(2n) =
\left(\frac{2^{2n}-2}{2^{2n}}\right) \xi(2n) \Gamma(2n),
\label{int-zerotemp-factor}
\end{multline}
where \(\Gamma(n)\) --- Gamma function, \(\xi(n)\) --- Riemann Xi function:
\begin{gather}
\Gamma(n) = \int \limits_0^\infty z^{n-1} e^{-z} dz = (n-1)!, \\
\xi(n) = \sum \limits_{m=1}^\infty \frac{1}{m^n}.
\end{gather}

Further, one can rewrite~(\ref{int-zerotemp-factor}) in terms of Bernoulli numbers. It can be shown, that for integer~\(n\):
\begin{equation}
\xi(2n) = (-1)^{n+1} (2\pi)^{2n} \frac{B_{2n}}{2(2n)!}
\end{equation}
where \(B_n\) --- \(n\)-th Bernoulli number.

Thus, for integral~(\ref{int-zerotemp-factor}) we have:
\begin{multline}
\int \limits_{0}^\infty \frac{e^z}{(e^z+1)^2} z^{2n} dz =
\left(\frac{2^{2n}-2}{2^{2n}}\right) \left[
(-1)^{n+1} (2\pi)^{2n} \frac{B_{2n}}{2(2n)!} \right] (2n-1)! = \\
= \left(\frac{2^{2n-1}-1}{2n}\right) (-1)^{n+1} \pi^{2n} B_{2n}.
\end{multline}

First order approximation of integral~(\ref{int-inital}) for low temperatures is:
\begin{equation}
I = \int \limits_0^\infty F(E) n'_f (E,T) dE \approx
- F(\mu) - \frac{F''(\mu)(k_BT)^2}{8} \pi^2 B_2.
\label{first-order-temp}
\end{equation}

Applying obtained expression~(\ref{first-order-temp}) to calculation of chemical potential and conductivity one may get the following expressions:
\begin{equation}
N_e = 2 - \int \limits_{E_{\text{min}}}^{E_{\text{max}}} n'_f (E,T) K(E) dE \approx
2 - K(\mu) - \frac{K''(\mu)(k_BT)^2}{8} \pi^2 B_2.
\end{equation}
\begin{multline}
\sigma = \frac{e^2}{\pi \hbar} \int \frac{1}{W(E)} \frac{\partial n_f}{\partial k} dE =
\frac{e^2}{\pi \hbar} \int \frac{1}{W(E)} \frac{\partial n_f}{\partial E}
\frac{\partial E}{\partial k} dE = \frac{e^2}{\pi \hbar} \int \frac{1}{W(E)}
n'_f(E) \hbar \sqrt{\frac{2E}{m}} dE =\\
= \frac{e^2}{\pi} \sqrt{\frac{2}{m}} \int \frac{\sqrt{E}}{W(E)} n'_f(E) dE
\approx - \frac{e^2}{\pi} \sqrt{\frac{2}{m}} \left[ \frac{\sqrt{\mu}}{W(\mu)}
+ \left( \left. \frac{\sqrt{E}}{W(E)} \right)'' \right|_{E=\mu}
\frac{\pi^2 B_2}{8} (k_BT)^2 \right].
\end{multline}

\section{Numerical results and discussion}

\begin{figure}[p]
\begin{tabular}{cc}
\includegraphics[width=0.45\textwidth]{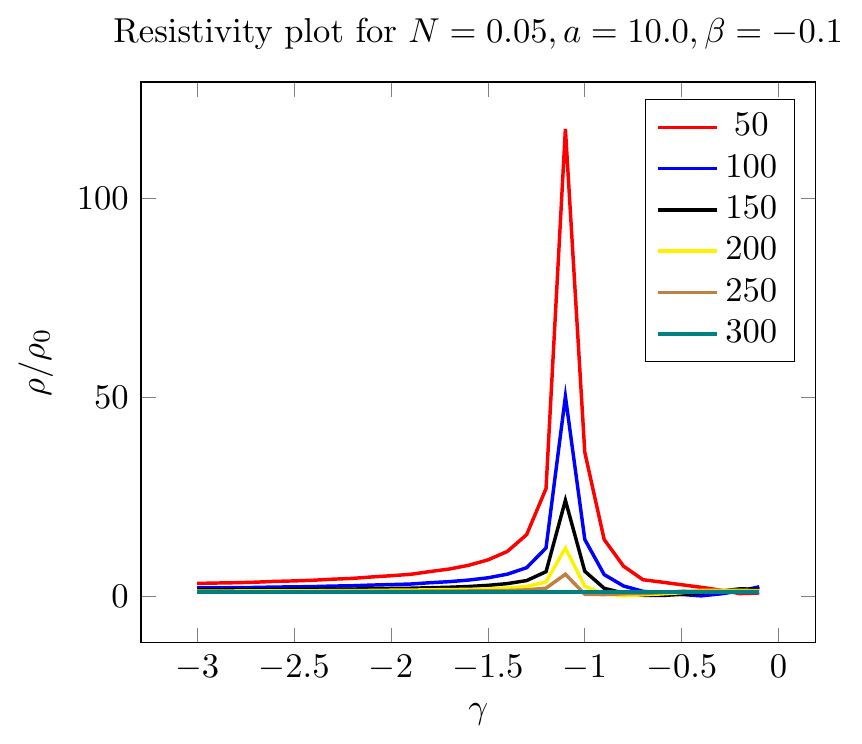} &
\includegraphics[width=0.45\textwidth]{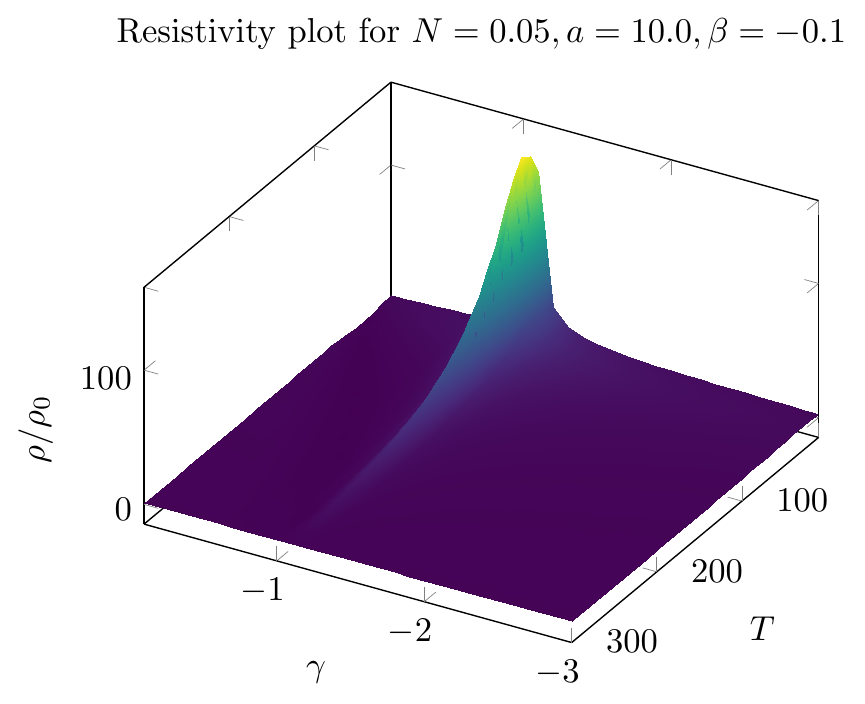} \\
\includegraphics[width=0.45\textwidth]{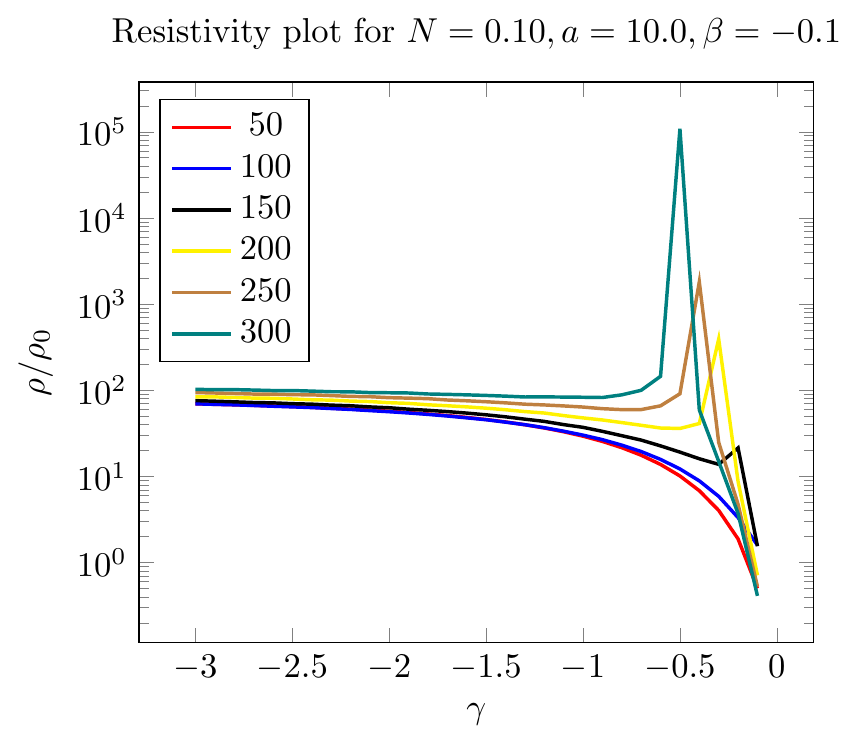} &
\includegraphics[width=0.45\textwidth]{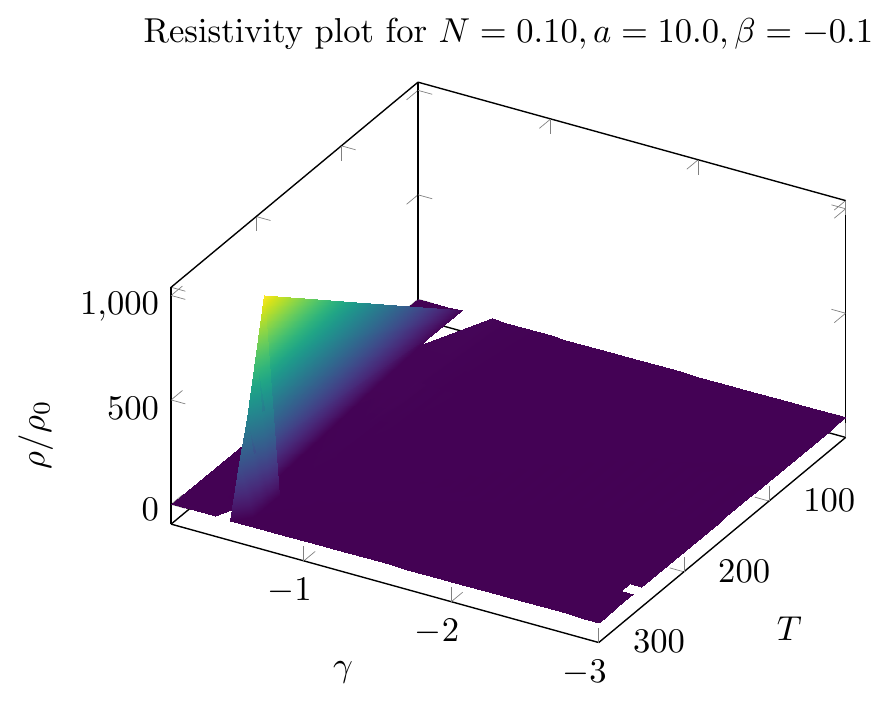} \\
\includegraphics[width=0.45\textwidth]{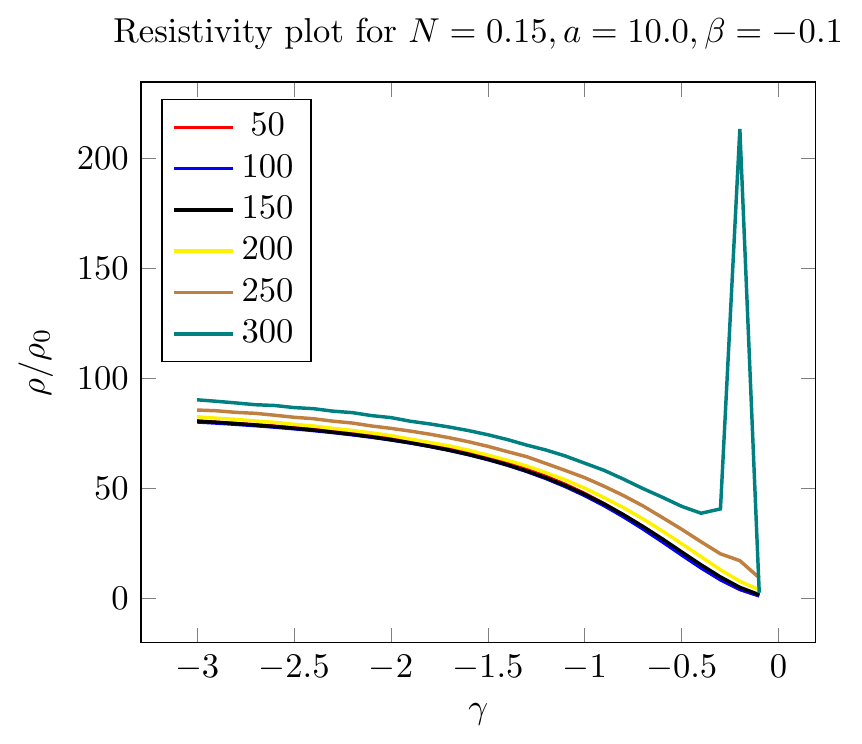} &
\includegraphics[width=0.45\textwidth]{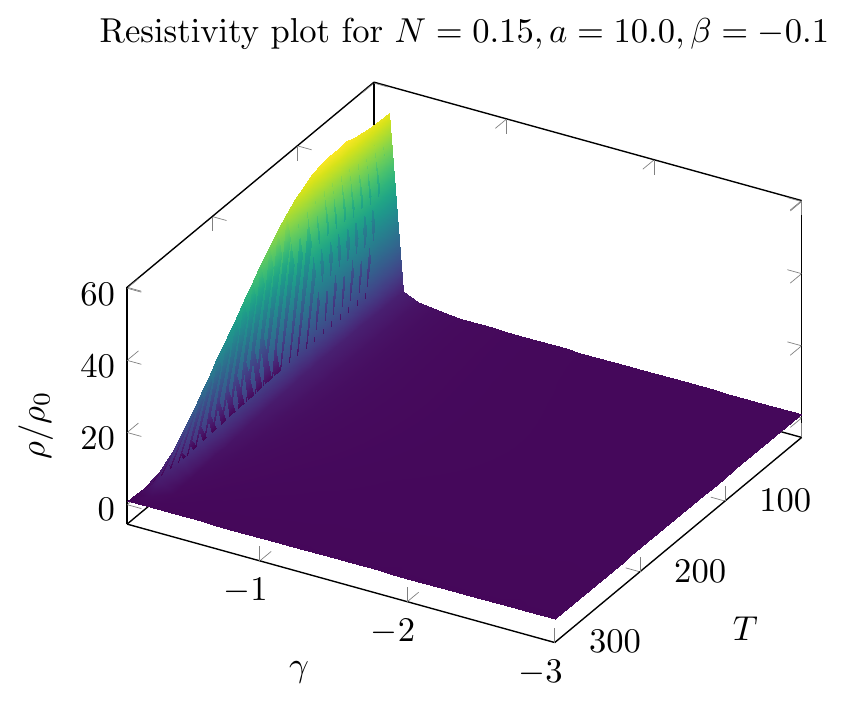} \\
\end{tabular}
\caption{Results of numerical experiments for conductivity (model parameters are: \(a=10\), \(\beta=-0.1\), \(x_0=a/2\))}
\label{num-exp}
\end{figure}

Some results of numerical experiments are shown in figure~(\ref{num-exp}).

In order to test the model, one have to substitute parameters according to Bi nanowires experimental data~\cite{Zhang00} and calculate conductivity plots.

For some \(\gamma\) values one can observe local extreme. It corresponds to the denominator of the~(\ref{resitivity-int}) approaches zero.

\section{Discussion}

Within given model conductivity plots were obtained. For certain model parameters \(\sigma/\sigma_0\) peak can be clearly seen. Thus, model can exhibit the behavior similar to that of the real systems. 
Since there is no clear way to construct averaging procedure, it is essential to choose model interpretation. There are two different ways of model implementation: a) choose model parameters in the way that certain basic physical quantities of real systems agrees with ones of model, b) fit model parameters to resemble experimental resistance behaviour.

It's known that for low temperatures (up to room temperatures) main contribution to a resistivity is made by a defect scattering. For a regular structure of quasi 1D nanoobject, surface can be treated as defect. Considering this, parameters radius dependence is as follows:
\( \beta=\text{const}, \quad \gamma=\gamma(r) \sim r\).
One over parameter, which depends on radius is full number of electrons (which is essential for Fermi energy and chemical potential calculation):
\( N_e = N_e (r) \sim \pi r^2 l\).

\section{Conclusions}
The model approbation showed promising results. The next step of the research will be two-dimensional model with cylindrical symmetry assumed.

\bibliography{references.bib}
\bibliographystyle{unsrt}

\end{document}